\RequirePackage{fix-cm}
\documentclass[smallextended]{svjour3}       
\smartqed  
\usepackage{graphicx}
\usepackage{graphicx}
\usepackage{algorithmic}
\usepackage{algorithm}
\usepackage{url}
\usepackage{longtable}
\usepackage{amsmath}
\usepackage{lineno}
\begin{document}
\linenumbers

\title{Solving the minimum sum coloring problem via binary quadratic programming
}


\author{Yang Wang \and Jin-Kao Hao* \and Fred Glover \and Zhipeng L\"u}


\institute{Yang Wang \at
              LERIA, Universit\'{e} d'Angers , 2 Boulevard Lavoisier, 49045 Angers Cedex 01, France \\
              \email{yangw@info.univ-angers.fr}           
           \and
           Jin-Kao Hao \emph{(Corresponding author)} \at
              LERIA, Universit\'{e} d'Angers , 2 Boulevard Lavoisier, 49045 Angers Cedex 01, France \\
              \email{hao@info.univ-angers.fr}
           \and
           Fred Glover \at
              OptTek Systems, Inc.,  2241 17th Street Boulder,CO 80302, USA \\
              \email{glover@opttek.com}
           \and
           Zhipeng L\"u \at
             School of Computer Science and Technology, Huazhong University of Science and Technology, 430074 Wuhan,
             China \\
              \email{zhipeng.lv@hust.edu.cn}
}

\date{\textbf{Submitted to Optimization Letters, May 4, 2012}}

\maketitle

\begin{abstract}
In recent years, binary quadratic programming (BQP) has been successively applied to solve several combinatorial
optimization problems. We consider in this paper a study of using the BQP model to solve the minimum sum coloring problem (MSCP). For this purpose, we recast the MSCP with a quadratic model which is then solved via a recently proposed Path Relinking (PR) algorithm designed for the general BQP. Based on a set of MSCP benchmark instances, we investigate the performance of this solution approach compared with existing methods.

\end{abstract}

\section{Introduction}
\label{Intro}

Given an undirected graph $G=(V,E)$ with vertex set $V$ and edge set
$E$, a K-coloring of $G$ is a function $c:v \mapsto c(v)$ that
assigns to each vertex $v \in V$ a color $c(v)$, where $c(v) \in \{1,
2, \ldots, K\}$. A K-coloring is considered legal if each pair of
vertices $(u,v)$ connected by an edge $(u,v) \in E$ receive
different colors $c(u)\neq c(v)$. The minimum sum coloring problem
(MSCP) is to find a legal K-coloring $c$ such that the total sum of
colors over all the vertices $\sum_{v \in V}{c(v)}$ is minimized.
The minimum value of this sum is called the chromatic sum of $G$ and
denoted by $\sum(G)$. The number of colors related to the chromatic
sum is called the strength of the Graph and denoted by $s(G)$.


The MSCP is NP-hard for general graphs \cite{Kubicka1989} and
provides applications mainly including VLSI design, scheduling and
resource allocation \cite{BarNoy1998,Malafiejski2004}. Given the
theoretical and practical significance of the MSCP, effective approximation
algorithms and polynomial algorithms have been presented for
some special cases of graphs, such as trees, interval graphs
and bipartite graphs
\cite{BN1998,Bonomo2011,Hajiabolhassan2000,Jansen2000,Kroon1996,Malafiejski2004,Salavatipour2003,Thomassen1989}.
For the purpose of practical solving of the general MSCP, a variety
of heuristics have been proposed in recent years, comprising a
parallel genetic algorithm \cite{Kokosinski2007}, a greedy algorithm
\cite{Li2009}, a tabu search algorithm \cite{Bouziri2010}, a hybrid
local search algorithm \cite{Douiri2011}, an independent set
extraction based algorithm \cite{Wu2012} and a local search
algorithm \cite{Helmar2011}.

On the other hand, binary quadratic programming (BQP) has
emerged during the past decade as a unified model for a wide range of combinatorial optimization problems, such as set packing \cite{Alidaee2008}, set partitioning \cite{Lewis2008}, generalized independent set \cite{Kochenberger2007},  maximum edge
weight clique \cite{Alidaee2007} and maxcut \cite{Wang2011b}. A review concerning the additional
applications and the reformulation procedures can be found in
\cite{Kochenberger2004}. This BQP approach has the advantage of
directly applying an algorithm designed for BQP to solve other
classes of problems rather than resorting to a specialized solution
method. Moreover, this approach proves to be competitive or even better than
the special algorithms proposed for several problems.

In this paper, we investigate for the first time the application of this BQP approach to solve the MSCP problem. We propose a binary quadratic formulation for the MSCP which is solved by our Path Relinking algorithm previously designed for the general BQP. To assess the performance of the proposed approach, we present computational results on a set of 23 benchmark instances from the literature and contrast these results with those of several reference algorithms specifically dedicated to the MSCP.

The rest of this paper is organized as follows. Section
\ref{sec_trans} illustrates how to transform the MSCP into the BQP
formulation. Section \ref{sec_pra} presents an overview of our Path
Relinking algorithm for the general BQP. Section
\ref{sec_results} is dedicated to computational results and
comparisons with other reference algorithms in the literature. The paper is concluded in Section \ref{sec_conclusion}.

\section{Transformation of the MSCP to the BQP model}
\label{sec_trans}

\subsection{Linear model for the MSCP}
\label{Linear model for the MSCP}

Given an undirected graph $G=(V,E)$ with vertex set $V$ ($n=|V|$)
and edge set $E$. Let $x_{uk}$ be 1 if vertex $u$ is assigned color
$k$, and 0 otherwise. The linear programming model for the MSCP can
be formulated as follows:
\begin{equation}\label{LP}
\begin{split}
& \ \ \ \ \ \ Min \ \ \ \ \ \ f(x) = \sum_{u=1}^{n}\sum_{k=1}^{K} k\cdot x_{uk} \\
&\text{subject to:  \ \ \ } c1. \ \ \sum_{k=1}^{K} x_{uk} = 1, \ \ u \in \{1,\ldots,n\} \\
& \ \ \ \ \ \ \ \ \ \ \ \ \ \ \ \ \ \ c2. \ \ x_{uk}+x_{vk} \leq 1, \ \forall(u,v)\in {E}, \ k \in \{1,\ldots,K\} \\
& \ \ \ \ \ \ \ \ \ \ \ \ \ \ \ \ \ \ c3. \ \ x_{uk}\in\{0,1\}
\end{split}
\end{equation}

\subsection{Nonlinear BQP alternative}
\label{Nonlinear BQP alternative}

The linear model of the MSCP can be recast into the form of the BQP
according to the following steps:

For the constraints $c1.$, we represent these linear equations by a matrix $Ax = b$ and incorporate the following penalty
transformation \cite{Kochenberger2004}:
\begin{equation}
\begin{split}
& \ \ \ \ \ \ \#1: \ \ \ \ \ \ f1(x)=P(Ax-b)^{t}(Ax-b) \ \ \ \ \ \ \ \ \ \ \ \ \ \ \ \ \ \ \ \ \ \ \ \ \ \ \ \ \ \ \ \ \ \ \\
& \ \ \ \ \ \ \ \ \ \ \ \ \ \ \ \ \ \ \ \ \ \ \ \ \ \ = P[x^{t}(A^{t}A)x-x^{t}(A^{t}b)-(b^{t}A)x+b^{t}b] \ \ \ \ \ \ \ \ \ \ \ \ \\
& \ \ \ \ \ \ \ \ \ \ \ \ \ \ \ \ \ \ \ \ \ \ \ \ \ \ = P[x^{t}(A^{t}A)x-x^{t}(A^{t}b)-(b^{t}A)x]+Pb^{t}b \\
& \ \ \ \ \ \ \ \ \ \ \ \ \ \ \ \ \ \ \ \ \ \ \ \ \ \ =xD_{1}x+c \ \
\ \ \ \ \
\end{split}
\end{equation}

For the constraints $c2.$, we utilize the quadratic penalty function
$g(x)=Px_{uk}x_{vk}$ to replace each inequality $x_{uk}+x_{vk} \leq 1$
in $c2.$ and add them up as follows \cite{Kochenberger2004}:
\begin{equation}
\begin{split}
& \ \ \ \ \ \ \#2: \ \ \ \ \ \ \ f2(x)=\sum_{u=1}^{n}\sum_{v=1,u \neq v}^{n} \sum_{k=1}^{K}w_{uv}x_{uk}x_{vk} \ \ \ \ \ \ \ \ \ \ \ \ \ \ \ \ \ \ \ \ \ \ \ \ \ \ \ \ \ \ \ \ \ \ \\
& \ \ \ \ \ \ \ \ \ \ \ \ \ \ \ \ \ \ \ \ \ \ \ \ \ \ \ =xD_{2}x \ \
\ \ \ \ \
\end{split}
\end{equation}
where $w_{uv}=P$ if $(u,v)\in {E}$ and 0 otherwise.

To construct the nonlinear BQP formulation $h(x)$, we first inverse
the minimum objective of the MSCP to be $-f(x)$ in accordance with
the general BQP model under a maximum objective, which becomes the
first component of $h(x)$. Then we add the penalty function $f1(x)$
into $h(x)$ such that $f1(x)=0$ if all the linear equations in $c1.$
are satisfied and otherwise $f1(x)$ is a penalty term with large
negative values. In the same way, we add the penalty function
$f2(x)$ into $h(x)$. Hence, the resulting BQP formulation for the
MSCP can be expressed as follows:

\begin{equation}
\begin{split}
& \ \ \ \ \ \ \ \ BQP: \ \ Max \ h(x)=-f(x)+f1(x)+f2(x) \ \ \ \ \ \ \ \ \ \ \ \ \ \\
& \ \ \ \ \ \ \ \ \ \ \ \ \ \ \ \ \ \ \ \ \ \ \ \ \ \ \ \ \ \ \ \ \
=xQ'x+c \ \ \ \ \ \ \
\end{split}
\end{equation}

Once the optimal objective value for this BQP formulation is
obtained, the minimum sum coloring value can be readily obtained by
taking its inverse value.

Further, a penalty scalar $P$ is considered to be suitable as long
as its absolute value $|P|$ is larger than half of the maximum color
($|P|>K/2$). Consider that penalty functions should be negative
under the case of a maximal objective, we select $P=-500$ for the 
benchmark instances experimented in this paper. The optimized
solution $x$ obtained by solving the nonlinear BQP formulation
indicates that such selection ensures both $f1(x)$ and $f2(x)$ equal
to 0. In other words, each variable $x_{uk}$ with the assignment of
1 in the optimized solution $x$ forms a feasible K-coloring in which
vertex $u$ gets the color $k$.

\subsection{An example of the transformation}

To illustrate the transformation from the MSCP to the BQP formulation, we consider the following graph with $|V|=4$ and expect to find a legal $K$-coloring with $K=2$.
\begin{figure}[!hbp]
  \centering\scalebox{0.40}{\includegraphics{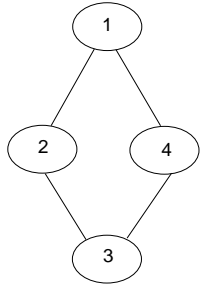}}\\
  \centering\caption{A small graph sample}\label{fig_mc2}
\end{figure}

Its linear formulation according to Equation (\ref{LP}) is:
\begin{equation} \nonumber
\begin{split}
& Max \ \ \ \ \ \ f(x)=-x_{11}-2x_{12}-x_{21}-2x_{22}-x_{31}-2x_{32}-x_{41}-2x_{42} \\
& \text{\ \ \ \ \ \ \ \ \ \ \ \ \ c1. \ \ \ \ \ \ \ \ } x_{11}+x_{12}=1;\\
& \ \ \ \ \ \ \ \ \ \ \ \ \ \ \ \  \ \ \ \ \ \ \ \ \ \ x_{21}+x_{22}=1; \\
& \ \ \ \ \ \ \ \ \ \ \ \ \ \ \ \  \ \ \ \ \ \ \ \ \ \ x_{31}+x_{32}=1;\\
& \ \ \ \ \ \ \ \ \ \ \ \ \ \ \ \  \ \ \ \ \ \ \ \ \ \ x_{41}+x_{42}=1; \\
& \text{\ \ \ \ \ \ \ \ \ \ \ \ \ c2. \ \ \ \ \ \ \ \ } x_{11}+x_{21} \leq 1; \ \ \ \ x_{12}+x_{22} \leq 1;\\
& \ \ \ \ \ \ \ \ \ \ \ \ \ \ \ \  \ \ \ \ \ \ \ \ \ \ x_{11}+x_{41} \leq 1; \ \ \ \ x_{12}+x_{42} \leq 1;\\
& \ \ \ \ \ \ \ \ \ \ \ \ \ \ \ \  \ \ \ \ \ \ \ \ \ \ x_{21}+x_{31} \leq 1; \ \ \ \ x_{22}+x_{32} \leq 1;\\
& \ \ \ \ \ \ \ \ \ \ \ \ \ \ \ \  \ \ \ \ \ \ \ \ \ \ x_{31}+x_{41} \leq 1; \ \ \ \ x_{32}+x_{42} \leq 1;\\
& \text{\ \ \ \ \ \ \ \ \ \ \ \ \ c3. \ \ \ \ \ \ \ \ } x_{11}, x_{12}, \ldots, x_{42} \in \{0,1\}\\
\end{split}
\end{equation}

Choosing the scalar penalty $P=-5$, we obtain the following BQP model:
\begin{equation} \nonumber
\begin{split}
& Max \ \ \ \ \ \ f(x)=xQx-20 \ \ \ \ \ \ \ \ \ \ \ \ \ \ \ \ \ \ \
\ \ \ \ \ \ \ \ \ \ \ \ \ \ \ \ \ \ \ \ \ \ \ \ \ \ \ \ \ \ \ \ \ \ \ \ \ \ \\
\end{split}
\end{equation}

where $xQx$ is written as:

\begin{equation} \nonumber
\left(
\begin{array}{c}
x_{11} \ x_{12} \ \ldots \ x_{41} \ x_{42}
\end{array}
\right) \times \left(
\begin{array}{cccccccc}
4 \ &-5 \ &-5 \ &0 \ &0 \ &0 \ &-5 \ &0 \\
-5 \ &3 \ &0 \ &-5 \ &0 \ &0 \ &0 \ &-5 \\
-5 \ &0 \ &4 \ &-5 \ &-5 \ &0 \ &0 \ &0 \\
0 \ &-5 \ &-5 \ &3 \ &0 \ &-5 \ &0 \ &0 \\
0 \ &0 \ &-5 \ &0 \ &4 \ &-5 \ &-5 \ &0 \\
0 \ &0 \ &0 \ &-5 \ &-5 \ &3 \ &0 \ &-5 \\
-5 \ &0 \ &0 \ &0 \ &-5 \ &0 \ &4 \ &-5 \\
0 \ &-5 \ &0 \ &0 \ &0 \ &-5 \ &-5 \ &3
\end{array}
\right) \times \left(
\begin{array}{c}
x_{11} \\x_{12}\\ \vdots \\ x_{41} \\ x_{42}
\end{array}
\right)
\end{equation}

Solving this BQP model yields $x_{11}=x_{22}=x_{31}=x_{42}=1$ (all other variables equal zero) and the optimal objective function value $f(x)=-6$. Reversing this objective function value leads to the optimum (the minimum sum coloring) of 6 for this graph.

\section{Path Relinking algorithm}
\label{sec_pra}

Path Relinking \cite{Glover2003} is a general search
strategy closely associated with tabu search and its underlying
ideas share a significant intersection with the tabu search
perspective \cite{Glover1989}, with applications in a variety of
contexts where it has proved to be very effective in solving
difficult problems. Our previous Path Relinking algorithm (PR) is
directly utilized to solve the MSCP in its nonlinear BQP form as
expressed in Eq. 4. The proposed Path Relinking algorithm is mainly
composed of the following components: \textit{RefSet
Initialization}, \textit{Solution Improvement}, \textit{Relinking},
\textit{Solution Selection}, \textit{RefSet Updating} and
\textit{RefSet Rebuilding}. Below we overview the general scheme of
our Path Relinking algorithm. More details can be found in
\cite{Wang2012}.

Firstly, the \textit{RefSet Initialization} method is used to create an initial set of elite solutions \textit{RefSet} where each
solution is obtained by applying a tabu search procedure to a randomly generated solution.

Secondly, for each pair of solutions ($x^{i}$, $x^{j}$) in \textit{RefSet} we undertake the following operations: (1) apply a \textit{Relinking} method to generate two paths (from $x^{i}$ to $x^{j}$ and from $x^{j}$ to $x^{i}$) by exploring trajectories (strictly confined to the neighborhood space) that connect high-quality solutions. The first and last solutions of the path are respectively called the initiating and guiding solutions. At each step of building the path from the initiating solution to the guiding solution, we randomly select a variable from the set of variables for which $x^{i}$ and $x^{j}$) have different values; (2) apply a \textit{Solution Selection} method to select one solution from those generated on the path by reference both to its quality and to the hamming distance of this solution to the initiating solution and the guiding solution. This selected solution is then submitted to the tabu search based \textit{Improvement Method}; (3) apply a \textit{RefSet Updating} method to decide if the newly improved solution is inserted in \textit{RefSet} to replace the worst solution. The above procedure is called a round of our Path Relinking procedure.

Finally, once a round of Path Relinking procedure is completed, a \textit{RefSet Rebuilding} method is launched to rebuild \textit{RefSet} that is used by the next round of the Path Relinking procedure. The Path Relinking algorithm terminates as soon as its stop condition (e.g. a fixed computing time) is satisfied. 

\section{Experimental results}
\label{sec_results}

\subsection{Experimental protocols}

To assess this BQP approach for the MSCP, we carry out experiments on a set of 23 graphs\footnote{http://mat.gsia.cmu.edu/COLOR/instances.html}, which are the most used benchmark instances in the literature. Our experiments are conducted on a PC with Pentium 2.83GHz CPU and 8GB RAM. The time limit for a single run of our Path Relinking algorithm is set as follows: 1 hour for the first 16 instances in Table
\ref{table_results}; 10 hours for dsjc125.1, dsjc125.5, dsjc125.9,
dsjc250.1 and dsjc250.5; 20 hours for dsjc250.9 and dsjc500.1. Given
the stochastic nature of our PR algorithm, each problem instance is
independently solved 20 times.

The tabu tenure $ttl$ and the improvement cutoff $\mu$
\cite{Wang2012} are two parameters in the tabu search based improvement
method--a key component of the PR algorithm. According to preliminary
experiments, we set $ttl=max\{40, N/100+rand(50)\}$ (where $N$
denotes the number of variables in the resulting BQP model and
rand(50) receives a random integer ranging from 1 to 50). In
addition, we set $\mu=2N$ for the improvement of the initial solutions in $RefSet$ and $\mu=500$ for the improvement of the
solutions on the path, respectively.

\subsection{Results of the BQP model for the MSCP}

Table \ref{table_results} presents the computational statistics of
the BQP model for the MSCP. Columns 1 to 3 give the instance names
\textit{Instances} along with the vertex number $V$ and edge number
$E$ of the graphs. Columns 4 and 5 show the number of colors $K$ to
be used and the number of variables $N$ in the BQP formulation.
Column 6 summarizes the best known results \textit{BKR} from the
previous literature
\cite{Bouziri2010,Douiri2011,Helmar2011,Li2009,Kokosinski2007,Wu2012}.
The columns under the heading of BQP-PR report our results of the
BQP model solved by the PR algorithm: the best objective values
$Best$, the average objective values $Avr$, the standard deviation
$\sigma$, the average time $T_{b\_avr}$ (in seconds) to reach the
best objective value $Best$ over 20 runs, and the average time (in
seconds) $T_{AVR}$ consumed to reach the best objective value
obtained in each run. The last row shows the average performance
over a total of 23 tested instances in terms of the deviations of
the best and average solution values from the \textit{BKR}. Notice
that the results marked in bold in the $Best$ column indicate that
BQP-PR reaches the \textit{BKR} on these instances.

We also applied CPLEX V12.2 to the linear model (\ref{LP}) (Section
\ref{Linear model for the MSCP}) to solve these
MSCP instances. CPLEX was successful in finding an optimal solution
for 15 instances (marked with an asterisk), but terminated
abnormally for the remaining 8 instances due to excess requirements
of memory.


\renewcommand{\baselinestretch}{1.0}\large\normalsize
\begin{table}\centering\scriptsize
\caption{Computational statistics of the BQP-PR approach for the
MSCP} \label{table_results}
\begin{tabular}{p{1.35cm}p{0.45cm}p{0.6cm}p{0.35cm}p{0.55cm}p{0.45cm}p{0.1pt}p{0.55cm}p{0.7cm}p{0.55cm}p{0.6cm}p{0.6cm}p{0.00001cm}}
\hline
& & & & & & & \multicolumn{5}{c}{BQP-PR} &\\
\cline{8-12}
\centering{\raisebox{1.5ex}[0cm][0cm]{\textit{Instances}}}  &  \centering{\raisebox{1.5ex}[0cm][0cm]{$V$}} & \centering{\raisebox{1.5ex}[0cm][0cm]{$E$}} & \centering{\raisebox{1.5ex}[0cm][0cm]{$K$}} &  \centering{\raisebox{1.5ex}[0cm][0cm]{$N$}} & \centering{\raisebox{1.5ex}[0cm][0cm]{\textit{BKR}}} & & \centering{$Best$} & \centering{$Avr$} & \centering{$\sigma$} & \centering{$T_{b\_avr}$} & \centering{$T_{AVR}$} &\\
\hline
\centering{myciel3}  &  \centering{11} & \centering{20} & \centering{6} &  \centering{66} & \centering{$21^*$}  & & \centering{\textbf{21}} & \centering{21.0}& \centering{0.0} & \centering{$<1$} & \centering{$<1$}&\\
\centering{myciel4}  &  \centering{23} & \centering{71} & \centering{7} &  \centering{161} & \centering{$45^*$} & & \centering{\textbf{45}} & \centering{45.0}& \centering{0.0} & \centering{$<1$} & \centering{$<1$}&\\
\centering{myciel5}  &  \centering{47} & \centering{236} & \centering{8} &  \centering{376} & \centering{$93^*$}  & & \centering{\textbf{93}} & \centering{93.0}& \centering{0.0} & \centering{2} & \centering{2}&\\
\centering{myciel6}  &  \centering{95} & \centering{755} & \centering{10} &  \centering{950} & \centering{$189^*$}  & & \centering{\textbf{189}} & \centering{189.0}& \centering{0.0} & \centering{497} & \centering{497}&\\
\centering{myciel7}  &  \centering{191} & \centering{2360} & \centering{10} &  \centering{1910} & \centering{381}  & & \centering{\textbf{381}} & \centering{384.9}& \centering{3.7} & \centering{70} & \centering{384}&\\
\centering{anna}  &  \centering{138} & \centering{493} & \centering{13} &  \centering{1794} & \centering{$276^*$}  & & \centering{\textbf{276}} & \centering{276.3}& \centering{0.4} & \centering{870} & \centering{721}&\\
\centering{david}  &  \centering{87} & \centering{406} & \centering{13} &  \centering{1131} & \centering{$237^*$}  & & \centering{\textbf{237}} & \centering{237.0}& \centering{0.0} & \centering{524} & \centering{524}&\\
\centering{huck}  &  \centering{74} & \centering{301} & \centering{13} &  \centering{962} & \centering{$243^*$} & &  \centering{\textbf{243}} & \centering{243.0}& \centering{0.0} & \centering{3} & \centering{3}&\\
\centering{jean}  &  \centering{80} & \centering{254} & \centering{12} &  \centering{960} & \centering{$217^*$}  & & \centering{\textbf{217}} & \centering{217.0}& \centering{0.0} & \centering{79} & \centering{79}&\\
\centering{queen5.5}  &  \centering{25} & \centering{160} & \centering{7} &  \centering{175} & \centering{$75^*$}  & & \centering{\textbf{75}} & \centering{75.0}& \centering{0.0} & \centering{$<1$} & \centering{$<1$}&\\
\centering{queen6.6}  &  \centering{36} & \centering{290} & \centering{9} &  \centering{324} & \centering{$138^*$}  & & \centering{\textbf{138}} & \centering{138.0}& \centering{0.0} & \centering{6} & \centering{6}&\\
\centering{queen7.7}  &  \centering{49} & \centering{476} & \centering{9} &  \centering{441} & \centering{$196^*$}  & & \centering{\textbf{196}} & \centering{196.0}& \centering{0.0} & \centering{6} & \centering{6}&\\
\centering{queen8.8}  &  \centering{64} & \centering{728} & \centering{10} &  \centering{640} & \centering{$291^*$}  & & \centering{\textbf{291}} & \centering{298.1}& \centering{5.1} & \centering{1064} & \centering{780}&\\
\centering{games120}  &  \centering{120} & \centering{638} & \centering{10} &  \centering{1200} & \centering{$443^*$}  && \centering{\textbf{443}} & \centering{446.5}& \centering{3.2} & \centering{755} & \centering{880}&\\
\centering{miles250}  &  \centering{128} & \centering{387} & \centering{10} &  \centering{1280} & \centering{$325^*$}  & & \centering{\textbf{325}} & \centering{328.6}& \centering{2.3} & \centering{777} & \centering{1704}&\\
\centering{miles500}  &  \centering{128} & \centering{1170} & \centering{22} &  \centering{2816} & \centering{$705^*$}  & & \centering{713} & \centering{722.5}& \centering{6.4} & \centering{653} & \centering{1942}&\\
\centering{DSJC125.1}  &  \centering{125} & \centering{736} & \centering{8} &  \centering{1000} & \centering{326}  & & \centering{329} & \centering{338.5}& \centering{6.3} & \centering{1684} & \centering{7115}&\\
\centering{DSJC125.5}  &  \centering{125} & \centering{3891} & \centering{22} &  \centering{2750} & \centering{1015}  & & \centering{1050} & \centering{1082.6}& \centering{20.2} & \centering{32924} & \centering{15186}&\\
\centering{DSJC125.9}  &  \centering{125} & \centering{6961} & \centering{50} &  \centering{6250} & \centering{2511}  & & \centering{2529} & \centering{2573.8}& \centering{26.2} & \centering{34801} & \centering{24650}&\\
\centering{DSJC250.1}  &  \centering{250} & \centering{3218} & \centering{12} &  \centering{3000} & \centering{977}  & & \centering{1027} & \centering{1062.9}& \centering{16.8} & \centering{8893} & \centering{17206}&\\
\centering{DSJC250.5}  &  \centering{250} & \centering{15668} & \centering{35} &  \centering{8750} & \centering{3246} & & \centering{3604} & \centering{3724.9}& \centering{59.1} & \centering{27009} & \centering{26065}&\\
\centering{DSJC250.9}  &  \centering{250} & \centering{27897} & \centering{80} &  \centering{20000} & \centering{8286} & & \centering{8604} & \centering{8869.4}& \centering{122.2} & \centering{70737} & \centering{65673}&\\
\centering{DSJC500.1}  &  \centering{500} & \centering{12458} & \centering{16} &  \centering{8000} & \centering{2850} & & \centering{3152} & \centering{3234.1}& \centering{41.7} & \centering{42447} & \centering{59241}&\\
\hline
\centering{Average}    & \centering{}         & \centering{}   &\centering{}       & \centering{}  & \centering{} & & \centering{0.0160} &\centering{0.0283} &\centering{13.63}    & \centering{9730.5}         & \centering{9681.1} & \\
\hline
\end{tabular}
\end{table}
\renewcommand{\baselinestretch}{1.0}\large\normalsize

From Table \ref{table_results}, we observe that our BQP-PR approach
is able to reach the best known results for 15 out of 23 instances,
among which 14 results are known to be optimal values. The table discloses that only the few 
best algorithms that are specifically tailored to the MSCP can compete with this performance (see also
Section \ref{subsec_C1}, Table \ref{table_cmp}). Moreover, the best
and average solution values obtained by BQP-PR are very close to the
best known results, with average deviations of 0.0160 and 0.0283,
respectively over the set of the benchmark instances.

A further observation is that our BQP-PR approach is quite robust to
reach optimal or best known solution values within a short period of
time for the instances with $N<2000$ variables in comparison with
a slow convergence for instances with many more BQP variables. We will further discuss this point in Section
\ref{sec_discussion}.


\subsection{Comparison with other special purpose algorithms for the MSCP}
\label{subsec_C1}

In order to further evaluate our BQP-PR approach, we show a comparison
of the proposed approach with several special purpose algorithms for the MSCP. These
algorithms include a hybrid local search algorithm HLS
\cite{Douiri2011}, an advanced recursive largest first algorithm
MRLF \cite{Li2009}, a parallel genetic algorithm PGA
\cite{Kokosinski2007}, a tabu search algorithm TS
\cite{Bouziri2010}, a very recent independent set extraction based heuristic EXSCOL \cite{Wu2012} and a recent local search heuristic MSD(5)
\cite{Helmar2011}. Given that the reference algorithms either utilize different termination conditions or do not report the
computing time, we base the comparison on solution quality. Table \ref{table_cmp} presents the best objective values
obtained by each algorithm (BQP-PR, HLS, MRLF, PGA, TS, EXSCOL
and MSD(5), respectively) where the best solution values among them
are marked in bold. The results for HLS, PGA and TS which are unavailable are marked with "--".

\renewcommand{\baselinestretch}{1.0}\large\normalsize
\begin{table}\centering\scriptsize
\caption{Comparison between the BQP-PR approach and other specific MSCP algorithms}
\label{table_cmp}
\begin{tabular}{p{1.35cm}p{0.55cm}p{1.1cm}p{0.55cm}p{1.05cm}p{0.9cm}p{0.4cm}p{1.25cm}p{1.25cm}p{0.0001cm}}
\hline
\centering{\textit{Instances}} &\centering{\textit{BKR}} & \centering{BQP-PR} & \centering{HLS\cite{Douiri2011}} & \centering{MRLF\cite{Li2009}} & \centering{PGA\cite{Kokosinski2007}} & \centering{TS\cite{Bouziri2010}} & \centering{EXSCOL\cite{Wu2012}} & \centering{MDS(5)\cite{Helmar2011}} & \\
\hline
\centering{myciel3}  &  \centering{$21^*$} &  \centering{\textbf{21}} & \centering{\textbf{21}}  & \centering{\textbf{21}} & \centering{\textbf{21}} & \centering{--} & \centering{\textbf{21}} & \centering{\textbf{21}}&\\
\centering{myciel4}  &  \centering{$45^*$} &  \centering{\textbf{45}} & \centering{\textbf{45}} & \centering{\textbf{45}} &  \centering{\textbf{45}} & \centering{--} & \centering{\textbf{45}} & \centering{\textbf{45}}&\\
\centering{myciel5}  &  \centering{$93^*$} &  \centering{\textbf{93}} & \centering{\textbf{93}} & \centering{\textbf{93}} &  \centering{\textbf{93}} & \centering{--} & \centering{\textbf{93}} & \centering{\textbf{93}}&\\
\centering{myciel6}  &  \centering{$189^*$} &  \centering{\textbf{189}} & \centering{\textbf{189}} & \centering{\textbf{189}} &  \centering{\textbf{189}} & \centering{--} & \centering{\textbf{189}} & \centering{\textbf{189}}&\\
\centering{myciel7}  &  \centering{$381$} &  \centering{\textbf{381}} & \centering{\textbf{381}} & \centering{\textbf{381}} &  \centering{382} & \centering{--} & \centering{\textbf{381}} & \centering{\textbf{381}}&\\
\centering{anna}  & \centering{$276^*$} &  \centering{\textbf{276}} & \centering{--} & \centering{277} &  \centering{281} & \centering{--} & \centering{283} & \centering{\textbf{276}} &\\
\centering{david}  &  \centering{$237^*$} &  \centering{\textbf{237}} & \centering{--}  & \centering{241} & \centering{243} & \centering{--} & \centering{\textbf{237}} & \centering{\textbf{237}}&\\
\centering{huck}  &  \centering{$243^*$} &  \centering{\textbf{243}} & \centering{\textbf{243}} & \centering{244} &  \centering{\textbf{243}} & \centering{--} & \centering{\textbf{243}} & \centering{\textbf{243}}&\\
\centering{jean}  &  \centering{$217^*$} &  \centering{\textbf{217}} & \centering{--} & \centering{\textbf{217}} &  \centering{218} & \centering{--} & \centering{\textbf{217}} & \centering{\textbf{217}}&\\
\centering{queen5.5}  &  \centering{$75^*$} &  \centering{\textbf{75}} & \centering{--} & \centering{\textbf{75}} &  \centering{\textbf{75}} & \centering{--} & \centering{\textbf{75}} & \centering{\textbf{75}}&\\
\centering{queen6.6}  &  \centering{$138^*$} &  \centering{\textbf{138}} & \centering{\textbf{138}} & \centering{\textbf{138}} &  \centering{\textbf{138}} & \centering{--} & \centering{150} & \centering{\textbf{138}}&\\
\centering{queen7.7}  &  \centering{$196^*$} &  \centering{\textbf{196}} & \centering{--} & \centering{\textbf{196}} &  \centering{\textbf{196}} & \centering{--} & \centering{\textbf{196}} & \centering{\textbf{196}} &\\
\centering{queen8.8}  &  \centering{$291^*$} &  \centering{\textbf{291}} & \centering{--}  & \centering{303} & \centering{302} & \centering{--} & \centering{\textbf{291}} & \centering{\textbf{291}}&\\
\centering{games120}  &  \centering{$443^*$} &  \centering{\textbf{443}} & \centering{446} & \centering{446} &  \centering{460} & \centering{--} & \centering{\textbf{443}} & \centering{\textbf{443}}&\\
\centering{miles250}  &  \centering{$325^*$} &  \centering{\textbf{325}} & \centering{343} & \centering{334} &  \centering{347} & \centering{--} & \centering{328} & \centering{\textbf{325}}&\\
\centering{miles500}  &  \centering{$705^*$} &  \centering{713} & \centering{755} & \centering{715} &  \centering{762} & \centering{--} & \centering{\textbf{709}} & \centering{712}&\\
\centering{DSJC125.1}  &  \centering{326} &  \centering{329} & \centering{--} & \centering{352} &  \centering{--} & \centering{344} & \centering{\textbf{326}} & \centering{\textbf{326}}&\\
\centering{DSJC125.5}  &  \centering{1015} &  \centering{1050} & \centering{--} & \centering{1141} &  \centering{--} & \centering{1103} & \centering{1017} & \centering{\textbf{1015}}&\\
\centering{DSJC125.9}  &  \centering{2511} &  \centering{2529} & \centering{--} & \centering{2653} &  \centering{--} & \centering{2631} & \centering{2512} & \centering{\textbf{2511}}&\\
\centering{DSJC250.1}  &  \centering{977} &  \centering{1027} & \centering{--} & \centering{1068} &  \centering{--} & \centering{1046} & \centering{985} & \centering{\textbf{977}}&\\
\centering{DSJC250.5}  &  \centering{3246} &  \centering{3604} & \centering{--} & \centering{3658} &  \centering{--} & \centering{3779} & \centering{\textbf{3246}} & \centering{3281}&\\
\centering{DSJC250.9}  &  \centering{8286} &  \centering{8604} & \centering{--} & \centering{8942} &  \centering{--} & \centering{9198} & \centering{\textbf{8286}} & \centering{8412}&\\
\centering{DSJC500.1}  &  \centering{2850} &  \centering{3152} & \centering{--} & \centering{3229} &  \centering{--} & \centering{3205} & \centering{\textbf{2850}} & \centering{2951}&\\
\hline
\end{tabular}
\end{table}
\renewcommand{\baselinestretch}{1.0}\large\normalsize

As we can observe in Table \ref{table_cmp}, the proposed BQP-PR approach outperforms HLS, MRLF,
PGA and TS in terms of the best solution values. Specifically, BQP-PR finds better solutions than HLS, MRLF, PGA and TS for 3, 14,
8 and 7 instances, respectively. Finally, BQP-PR performs less well compared with the most effective MSCP heuristics EXSCOL and MDS(5). From this experiment, we conclude that our BQP approach combined with the PR algorithm for tacking the MSCP
constitutes an interesting alternative to specific algorithms tailored to this problem.

\subsection{Discussion}
\label{sec_discussion}

In this section, we discuss some limitations of the proposed BQP-PR approach for solving the MSCP. First, the proposed method may require considerable computing time to reach its best solutions for large graph instances (see column $T_{AVR}$ in Table \ref{table_results}). This can be partially explained by the fact that the number of the  BQP variables (equaling $V \cdot K$ where $V$ is the number of vertices and $K$ is the number of colors) sharply increases with the growth of $V$ and $K$. Additionally, at present our approach is not able to solve graph
instances with BQP variables surpassing the threshold value of 20,000 because of the memory limitation. These obstacles could be overcome by designing more effective data structures used by the BQP algorithms.

\section{Conclusion}
\label{sec_conclusion}

We have investigated the possibility of solving the
NP-hard minimum sum coloring problem (MSCP) via binary quadratic
programming (BQP). We have shown how the MSCP can be recast into the
BQP model and explained the key ideas of the Path-Relinking
algorithm designed for the general BQP. Experiments on a set of
benchmark instances demonstrate that this general BQP approach is
able to reach competitive solutions compared with several special
purpose MSCP algorithms even though considerable computing time may
be required.

\section*{Acknowledgment}

The work is partially supported by the ``Pays de la Loire" Region (France) within the RaDaPop (2009-2013) and LigeRO (2010-2013) projects.

\end{document}